\documentclass[12pt]{article}
\usepackage{amssymb}
\usepackage{graphicx}
\usepackage{amsmath}

\def\d{\partial}
\def\l{\left(}
\def\r{\right)}

\newcommand{\be}{\begin{equation}}
\newcommand{\ee}{\end{equation}}
\newcommand{\ba}{\begin{align}}
\newcommand{\ea}{\end{align}}
\newcommand{\bg}{\begin{gather}}
\newcommand{\eg}{\end{gather}}
\newcommand{\bseq}{\begin{subequations}}
\newcommand{\eseq}{\end{subequations}}

\def\half{\frac{1}{2}}

\begin{document}

\title{Free scalar dark matter candidates in $R^2$-inflation: \\
the light, the heavy and the superheavy}

\author{
D.~S.~Gorbunov\thanks{{\bf e-mail}: gorby@ms2.inr.ac.ru},
A.~G.~Panin\thanks{{\bf e-mail}: panin@ms2.inr.ac.ru}
\\
{\small{\em
Institute for Nuclear Research of the Russian Academy of Sciences,
}}\\
{\small{\em
60th October Anniversary prospect 7a, Moscow 117312, Russia
}}
}
\date{}

\maketitle

\begin{abstract}
Gravity takes care of both inflation and subsequent reheating in
Starobinsky's $R^2$-model. The latter is due to inflaton gravitation
decays dominated by scalar particle production. It is tempting to
suggest that dark matter particles are also produced in this
process. Since free scalars being 
too hot cannot serve as viable dark matter\,\cite{Gorbunov:2010bn},
we further study the issue considering two options: scalars with
non-minimal coupling to gravity and superheavy scalars generated at
inflationary stage. We found that the first option allows for viable
warm or cold dark matter if scalar mass exceeds $1.1$\,MeV.
The second option implies supercold dark
matter with particle mass $10^{16}$\,GeV,
which production is saturated at the end of inflation when
inflaton-dependent scalar mass rapidly changes and violates
adiabaticity. Similar result holds for superheavy fermion dark
matter.
\end{abstract}


Cosmology of the homogeneous and isotropic Universe in
Starobinsky's $R^2$-model\,\cite{starobinsky} has an inflationary
stage. This stage is realized as a large-field slow-roll inflation driven by
scalaron field, that is a scalar degree of freedom coming from gravity sector
of the theory. Subsequent decay of the scalarons into ordinary matter
dominated by scalar particles production reheats the
Universe. It is natural to speculate that dark matter particles could
be produced in the same way. 

Indeed, it is gravity---the mostly universal force---that provides interaction
for the scalaron with all (matter) fields. The strength of 
coupling to a given field is determined by the strength of conformal
symmetry violation. This leads to the fact that mass terms,
being conformally noninvariant, become modulated by the
scalaron field. Moreover, scalars have extra coupling due to
their conformally-noninvariant kinetic term. Consequently, the scalaron
decays are dominated by scalar particles
production\,\cite{starobinsky,Vilenkin:1985md}, so that
free scalars produced in an appropriate amount would be too hot and
could not serve as the dark matter~\cite{Gorbunov:2010bn}. At the same
time, free heavy fermions can do the job. However, scalar dark matter
implies only one new degree of freedom to be added to the SM, and
therefore is the most economic solution.

There are two obvious ways to get rid of the problem with scalar dark matter
and avoid over-energetic scalar relics.
First, one can partially restore the conformal symmetry
by adding nonminimal coupling of scalar to gravity.
By changing the value of corresponding coupling constant $\xi$ one can
control the strength of conformal symmetry violation and govern
the scalaron decay rate into dark matter particles. Then heavier and
hence colder scalar relics become viable dark matter.
The second option is to consider superheavy scalar particles
which kinematically can not be produced from scalaron decays
at postinflationary stage. However, some amount of them can be
generated nonperturbatively in the time-dependent scalaron
background.

In this work we consider both options. We obtain
the mass of the scalar dark matter candidate as a function of nonminimal
coupling to gravity $\xi$. Depending on the value of $\xi$
dark matter produced at postinflationary epoch can be light or heavy,
the lightest viable free scalars are of 1.1\,MeV and form warm dark matter.
Such particular candidate is interesting in the context of 
emerging problems of the standard CDM cosmology on small scales,
such as missing satellites, galactic matter density profiles and angular
momentum of spiral galaxies, for review see
e.g.\,\cite{Bertone:2004pz}.

Considering the second option we find that the production of
superheavy particles is saturated at the end of inflation, when
the scalar mass modulated by the scalaron field changes rapidly.
A nontrivial result is that the mass dependence of the number
density of created particles is a power law. Thereby in this model it is
possible to produce superheavy relics with mass greatly exceeding
the scalaron mass, which form supercold dark matter.

We start by recalling
some relevant for the present study facts about $R^2$-inflationary
model. The gravitational sector of this model is described by the
following Lagrangian
\be
\label{eq:1}
S^{JF}=-\frac{M_P^2}{2}\int \!\! \sqrt{-g}\,d^4x\,
\l R-\frac{R^2}{6\,\mu^2} \r\;,
\ee
where $R$ is scalar curvature, $\mu$ is dimensionful parameter, and
we introduce the reduced Planck mass $M_P =
M_{Pl}/\sqrt{8\pi}=2.4\times 10^{18}$~GeV.
Variation of the
action~\eqref{eq:1} with respect to $g_{\mu \nu}$ yields the fourth order
differential equation. One can argue (using general coordinate
invariance) that the degrees of freedom of the field $g_{\mu \nu}$ can
be split into massless spin-2 field $\tilde{g}_{\mu\nu}$ and
massive scalar field $\phi$ which obey the second order
equations of motion~\cite{Magnano}. This splitting is achieved
explicitly in the action written in the Einstein frame,
which one can come to by conformal transformation
\be
\label{eq:2}
g_{\mu\nu}\to \tilde{g}_{\mu\nu}=\chi\, g_{\mu\nu}\;, ~~~~~~
\chi= {\rm exp}\l \sqrt{2/3}\,\phi/M_P\r\;.
\ee
For the action~\eqref{eq:1} we have then
\be
\label{eq:3}
S^{EF}=\int \!\!\sqrt{-\tilde g}\,d^4x\, \left[
  -\frac{M_P^2}{2}\,\tilde R + \half\, \tilde{g}^{\mu\nu} \d_\mu \phi
  \d_\nu\phi -\frac{3\,\mu^2 M_P^2}{4}\, \l 1-\frac{1}{\chi\l \phi\r}
  \r^2 \right]\;, 
\ee 
everywhere $\tilde{g}_{\mu\nu}$ stands for the
metric tensor and $\tilde{R}$ is scalar curvature for the metric
$\tilde{g}_{\mu\nu}$.  One observed that in the Einstein frame the
action splits into the Einstein--Hilbert action for
$\tilde{g}_{\mu\nu}$ and ordinary action for the scalar field $\phi$ 
with a specific potential. The scalar mode in the action~\eqref{eq:1}
was named scalaron~\cite{starobinsky} and we will use this name for
the field $\phi$ in what follows.

The scalaron potential in~\eqref{eq:3} is exponentially flat at
super-Planckian field values and provides the slow-roll inflation.
Normalization of the amplitude of scalar perturbations,
generated during this stage,
to the observed CMB anisotropy requires~\cite{Faulkner:2006ub}
\[
\mu = 1.3\times 10^{-5}\; M_P\;.
\]
The spectral index and parameters of the generated tensor
perturbations (for recent discussion see\,\cite{Bezrukov:2011gp})
 are consistent with observational
constraints~\cite{Komatsu:2008hk}.

When the slow-roll conditions get violated, inflation terminates
and the inflaton $\phi$ starts to oscillate rapidly, with frequency
equal to scalaron mass $\mu$. This drives the Universe expansion
like at matter-dominated stage. The intermediate stage naturally ends
up with inflaton decays into ordinary particles. Here we briefly discuss
this process, see~\cite{starobinsky,Vilenkin:1985md,Gorbunov:2010bn}
for details.

The major role in the scalaron decay 
is played by scalars. Below we consider scalar $\varphi$ described by
the following Lagrangian in the Jordan frame
\be
\label{eq:4}
S^{JF}_{\varphi} = \int \!\! \sqrt{-g}\,d^4x\,
\l
\half \, g^{\mu\nu}\d_\mu\varphi\d_\nu\varphi
-\half \, m_\varphi^2\varphi^2 + \frac{\xi}{2} R \varphi^2
\r\;,
\ee
where the last term represents nonminimal coupling to gravity with
dimensionless parameter $\left|\xi\right|\lesssim 1$.
Note that with $\left|\xi\right|\lesssim 1$ we do not introduce a new
scale. Below we assume that $\xi$ is a free small parameter, but
let us single out two special points: {\it minimal coupling}
with $\xi = 0$ and {\it conformal coupling} corresponding to
$\xi = 1/6$. The first value leads to a free scalar field theory while
the second one ensures conformal invariance of the
action~\eqref{eq:4} without mass term. The mass term explicitly
breaks conformal invariance and one can expect (small)
deviation of $\xi$ from the conformal value due to quantum corrections.
Quantum corrections can also generate small but nonzero value
of $\xi$ for a scalar field minimally coupled to gravity.
Therefore one can expect that the value of coupling $\xi$
should be within some intervals covering $\xi=0$ or
$\xi=1/6$ rather than be exactly equal to $\xi=0$ or 
$\xi=1/6$, respectively. Below we are interested in only positive values 
of nonminimal coupling. Negative $\xi$ may lead to a nonadiabatic
evolution and thereby an 
amplification of scalar perturbations on superhorizon scales during 
inflation~\cite{Starobinsky:2001xq}. That in turn may change 
predictions for scalaron mass $\mu$ and requires detailed study 
beyond the scope of this paper.    

After conformal transformation~\eqref{eq:2} and field rescaling
\[
\varphi\to\tilde \varphi = \chi^{-1/2}\,\varphi\;
\]
the action~\eqref{eq:4} takes the form
\be
\label{eq:5}
\begin{split}
S^{EF}_{\varphi} =  \int \!\! \sqrt{-\tilde{g}}\,d^4x\,
& \left [
\half \, \tilde g^{\mu\nu}\d_\mu\tilde\varphi\d_\nu\tilde \varphi
+ \frac{\xi}{2} \, \tilde R \tilde \varphi^2
-\frac{1}{2 \chi}\, m_\varphi^2 \tilde \varphi^2 \right. \\
 + & \left. \half\, \l \frac{1}{6} - \xi \r \frac{\tilde \varphi^2}{M_P^2} \tilde g^{\mu \nu}
\d_\mu \phi \d_\nu \phi
+ \sqrt{6} \l \frac{1}{6} - \xi \r  \frac{\tilde \varphi}{M_P} \tilde g^{\mu \nu}
\d_\mu \tilde \varphi \d_\nu \phi\;
\right ]\;.
\end{split}
\ee
One can find a conformal transformation involving $\varphi$
and leading to another Einstein frame, where the nonminimal coupling
disappears (see, e.g.~\cite{Starobinsky:2001xq}).
This frame is more convenient to study the case of
two-field inflation, where the dynamics of scalar field $\varphi$ is
also relevant at the inflationary epoch. 
However, below we are interested in a more
simple situation, where only scalaron is responsible for 
the inflation, and adopt the metric~\eqref{eq:2}.

Equation\,\eqref{eq:5} describes the scalaron interaction with free
scalars. It  doesn't vanish at $\xi=0$ and
provides scalaron decay into a pair of scalars. The
second term in Eq.~\eqref{eq:5} yields on nontrivial cosmological
background a contribution to the scalar mass.
However, at reheating one has $|\tilde{R}| \ll \mu^2$, 
and the scalaron decay into scalars
of mass $m_\varphi<\mu/2$
remains kinematically
allowed\footnote{At the inflationary stage the situation is
different, as we discuss later on.}. Its width reads\,\cite{Vilenkin:1985md}
\be
\label{eq:6}
\Gamma_\xi = \l 1-6\xi +2 \frac{m_\varphi^2}{\mu^2}\ \r^2 \,
\frac{\mu^3\, \sqrt{1-4 m_\varphi^2/\mu^2}}{192 \pi M_P^2}\;.
\ee
Massless scalar field with $\xi = 1/6$ is conformally invariant and
does not interact with the scalaron, which is illustrated by
nullifying decay rate \eqref{eq:6} at $\xi \to 1/6$, $m_\varphi\to
0$. The mass term in Eq.~\eqref{eq:5} explicitly breaks conformal
invariance and the scalaron decay rate in this case is proportional to
$m_\varphi^4$, see Eq.~\eqref{eq:6}.  Hence, if the mass of the
conformally coupled scalars is very small, $m_\varphi \ll \mu$, the
scalaron decays very slowly.

The value of the coupling constant, which is natural
and guarantees the successful reheating through scalaron decays,
is $\xi =0$. In case of the Standard Model with four scalar
degrees of freedom (in
the Higgs sector) minimally coupled to gravity
scalaron rapidly decays to scalars thereby reheating the Universe upto the
temperature \cite{starobinsky,Vilenkin:1985md,Gorbunov:2010bn}
\be
\label{eq:7}
T_{reh}\approx 3.1\times 10^9~{\rm GeV}\;.
\ee



Now let us consider a new scalar field, which does not interact
with SM particles or this interaction is very weak, so these particles
never equilibrate in the primordial plasma. If stable at cosmological
time scale, such scalar is a good candidate to be dark matter.
Previously~\cite{Gorbunov:2010bn} we have shown that  minimally
coupled scalars are too hot to be dark matter. Here we
rehabilitate light scalars as a possible dark matter candidate
by introducing nonminimal interaction with gravity.

As we discussed above, light scalars are eventually produced from
scalaron decays. In order to describe this process we will treat the 
scalaron field configuration as a condensate of nonrelativistic 
particles with number density $n_{\phi}(t)$ and energy density 
$\rho_\phi(t) = \mu\,n_\phi(t)$. At some moment $t_*$ the number 
density of scalar particles produced from scalaron decays is
\[
dn(t_*) = 2\,\Gamma_\xi n_\phi(t_*)dt_* =
2\, \frac{\Gamma_\xi}{\mu}\rho_\phi(t_*)dt_*\;.
\]
To get the number density at reheating time $t_{reh}$ one should take
into account the volume expansion,
\be
\label{dn}
dn(t_{reh}) = dn(t_*)\frac{a^3(t_*)}{a^3(t_{reh})} =
2\, \frac{\Gamma_\xi}{\mu}\rho_\phi(t_*)\frac{a^3(t_*)}{a^3(t_{reh})} dt_*\;.
\ee
Now let us look at the physical momentum of these particles. At production
time $t_*$ it equals $p_* = \sqrt{\mu^2/4-m^2_\phi}$. Then it gets
redshifted and at reheating time one has
\be
\label{preh}
p_{reh} = p_*\frac{a(t_*)}{a(t_{reh})}\;.
\ee
Using this formula one can find the momentum difference at reheating
time for the particles produced from $t_*$ to $t_*+dt_*$,
\be
\label{dpreh}
dp_{reh} = p_{reh} H(t_*)dt_*\;.
\ee
On the other hand one can introduce the distribution function $f(p_{reh})$
of produced particles as $d n(t_{reh}) =
f(p_{reh})\,d^3p_{reh}/(2\pi)^3$. 
Substituting Eqs.~\eqref{dn},~\eqref{dpreh} into this formula, we get
\be
\label{dfunc}
f(p_{reh}) = 4 \pi^2\,\frac{\Gamma_\xi}{\mu}\,
\frac{\rho_\phi(t_*)}{p_{reh}^3 H(t_*)}\, \frac{a^3(t_*)}{a^3(t_{reh})}\;,
\ee
where production time $t_*$ should be expressed in terms of momentum
by Eq.~\eqref{preh}. 
At $t_* \ll t_{reh}$ one rewrites Eq.~\eqref{dfunc} as
\be
\label{funcsimpl}
f(p_{reh}) = 12 \pi^2\,\frac{\Gamma_\xi}{\mu} \frac{M_P^2 H(t_*)}{p_*^3}\;.
\ee

\begin{figure}
\centerline{\includegraphics[width=0.7\textwidth]{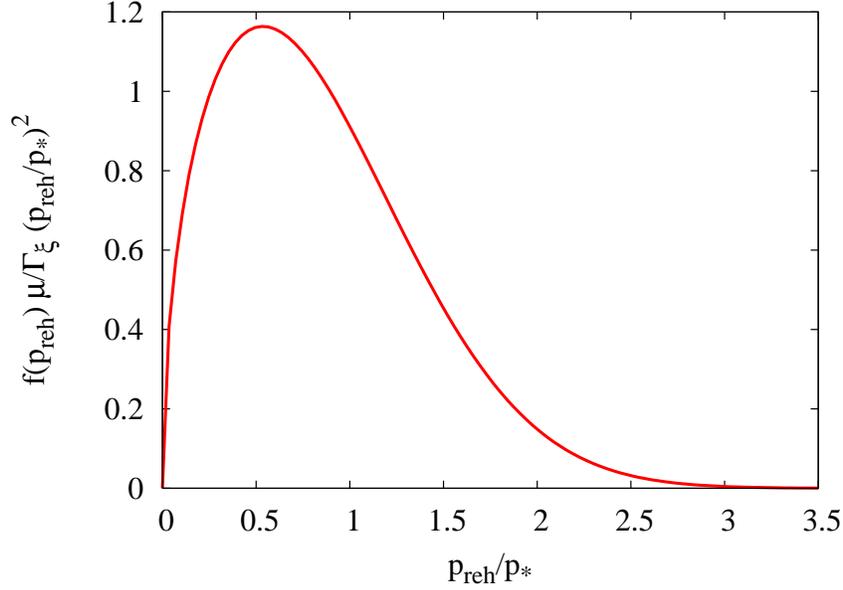}}
\caption{\label{spec} The distribution function for the scalars
produced from scalaron decays.}
\end{figure}

As one can see from Eq.~\eqref{dfunc}, for calculation of the spectrum 
one needs to know the time dependencies of the scale factor, Hubble 
parameter and scalaron energy density. In order to obtain a more 
accurate expressions, we use numerical calculations. Our result for 
the spectrum of produced scalars is presented in Fig.~\ref{spec}. 

For the particles number density one obtains then 
\be
\label{eq:11}
n_\varphi \simeq 2.5\,\frac{\Gamma_\xi}{\mu}  M_P^2 H_{reh} \;,
\ee
where numerical factor in front of r.h.s. of Eq.~\eqref{eq:11} is the 
result of numerical integration of the spectra in Fig.~\ref{spec} over 
momenta.
For the average momentum one gets
\be
\label{eq:12}
 \bar{p}_{reh} = 0.85\,p_*\;.
\ee
Note that $\bar{p}_{reh} \gg T_{reh}$, so the particles are very 
hot in comparison with the plasma.

In order to find the present abundance of the dark matter
one uses the ratio of the entropy to the number density.
At the reheating time we have
\be
\label{eq:13}
\frac{s}{n(t_{reh})} \simeq 0.2\,
\frac{\pi \sqrt{g_*} \mu T_{reh}}{\Gamma_\xi M_P}\;,
\ee
where $g_*$ is the effective number of degrees of freedom in the
plasma, see e.g.\,\cite{Gorbunov:2011zz}. 
The ratio \eqref{eq:13} remains intact during the subsequent hot stages
of the Universe expansion including the present epoch. From the
requirement that the relative contribution of the dark
matter to the present energy density $\rho_c$ is~\cite{Komatsu:2008hk}
$\Omega_{DM} = 0.223$ one gets the following relation
between the particle mass and nonminimal coupling $\xi$ which enters
Eq.~\eqref{eq:6},
\be
\label{eq:14}
m_\varphi \simeq 0.2\,
\frac{\Omega_{DM} \rho_c}{s_0}  
\frac{\pi \sqrt{g_*} \mu T_{reh}}{\Gamma_\xi M_P}\;.
\ee

Finally let us find the lower and the upper bounds on the scalar mass.
The lower bound corresponds to the case of warm dark matter. These particles
should have the average velocity $v^{max}_{eq} \sim 10^{-3}$ at the
epoch of equilibrium between radiation and matter densities.
On the other hand, the average velocity at that time can be found from
Eq.\,\eqref{eq:12}. That leads to the particle mass
\be
\label{eq:15}
m_{\varphi,\,min} \simeq \frac{0.42}{v^{max}_{eq}}\,\l 
\frac{g_{*,eq}}{g_*}\r^{\frac{1}{3}}
\frac{T_{eq}}{T_{reh}} \,  \mu \simeq 1.1~\text{MeV}\;,
\ee
where at the equilibrium $T_{eq}\approx0.76$\,eV, $g_{*,eq}\approx
3.9$ and $g_*=106.75$\,\cite{Gorbunov:2011zz}.  Then
Eq.\,\eqref{eq:14} implies $\xi = 1/6 \pm 0.018$ 
for the viable dark matter candidate.   

\begin{figure}[t]
\centerline{\includegraphics[width=0.7\textwidth]{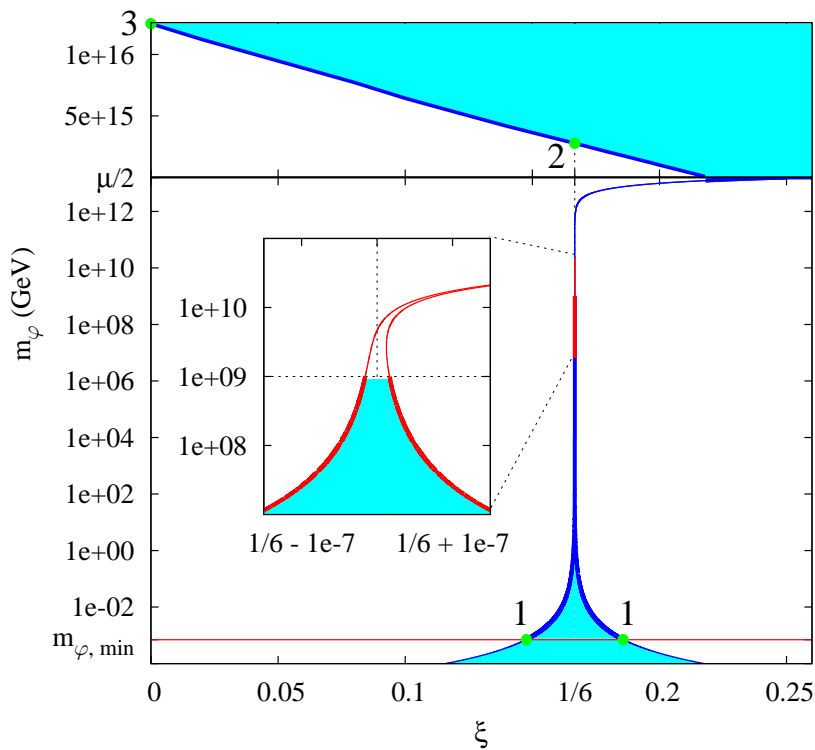}}
\caption{\label{fig:2} Mass of the scalar dark matter candidates
as a function of nonminimal coupling constant $\xi$ (solid thick
lines). The filled area represents cosmologically
allowed masses of free
scalars. At the bottom part of the plot the allowed
region is located between horizontal lines corresponded to
$m_\varphi \simeq 10^9$ GeV and $m_\varphi \simeq 1.1$ MeV.
The zoom shows the fine structure of the
curve near the conformal point $\xi = 1/6$.}
\end{figure}

The upper bound can be imposed if we take into account the
gravitational particle production from vacuum
fluctuations in the expanding Universe. It has been 
shown~\cite{Mamaev,Kuzmin:1998kk} that the particles conformally
coupled to gravity and heavier than $m_{\varphi,\,max} \simeq 10^9$ GeV
overclose the Universe. Possible dark matter masses~\eqref{eq:14}
together with the lower and upper bounds are shown in Fig~\ref{fig:2}.

The aforesaid consideration is valid while the occupation number for
produced particles remains small, $f(p_{reh})\lesssim 1$. Otherwise,
coherent effects become important and the corresponding occupation number
undergoes exponential growth. Let us check the maximum
occupation number, which in our case is reached by the particles
produced at the end of inflation as one can see from
Eq.~\eqref{funcsimpl}.
Using the relation~\eqref{eq:14} one finds that the maximal occupation number
exceeds unity only for the very light dark matter particles with 
\begin{equation}
\label{hot}
m_\varphi \lesssim  1~\text{keV}\;.
\end{equation}
These values of $m_\varphi$ are smaller than~\eqref{eq:15}, hence the coherent
effects become significant for the irrelevant hot dark matter only, and our
result~\eqref{eq:14} remains intact.


We proceed to
the case, when dark matter particle 
is heavier than the half-scalaron. Perturbative production
of such particles is kinematically forbidden. However, some amount of them
is eventually produced in a time-dependent scalaron background. In order to
describe this process one can use the method based on the Bogoliubov's
transformation coefficients~\cite{Kofman:1994rk}. Here we summarize
the basic formalism  which we employed in our analysis. For more
details see, e.g Ref.~\cite{Gorbunov:2011zzc}.

We start by canonically quantizing the action~\eqref{eq:5}  in curved
cosmological background with the external classical scalaron field.
For this purpose it's convenient to choose the FLRW metric in the form
$ds^2 = a^2(\eta) (d\eta^2 - d\vec{x}^2)$, where $a(\eta)$ is the scale
factor and $\eta$ is the conformal time defined as $d \eta = dt /a$.
After rescaling the field variable,
\[
\tilde{\varphi} = s/a(\eta)\;,
\]
the equation of motion following from the action~\eqref{eq:5} becomes
\be
\label{eq:16}
\left \{
\frac{\d^2}{\d\eta^2}  - \frac{\d^2}{\d\vec{x}^2}  + \frac{1}{\chi} \,a^2 m_\varphi^2
-\l \frac16 - \xi \r \l 6 \frac{a''}{a} + \frac{\phi'^2}{M_P^2}
+ \frac{\sqrt{6}\,a^2}{M_P} \frac{\d V(\phi)}{\d \phi} \r
\right \} s(\eta,\,\vec{x}) =0\;,
\ee
where prime means the derivative with respect to conformal time
and $V(\phi)$ is the scalaron potential, see Eq.~\eqref{eq:3}.
Here we have exploited the classical equation of motion to express the second
conformal time derivative of the scalaron field.
Solution of Eq.~\eqref{eq:16} can be written in the following form
\be
\label{eq:17}
s(\eta, \vec{x}) = \frac{1}{(2 \pi)^{3/2}} \int d^3p
\l
\hat{a}_p s_p(\eta) e^{-i \vec{p}\vec{x}} +
\hat{a}_p^\dag s_p^*(\eta) e ^{i \vec{p}\vec{x}}
\r\;,
\ee
where $\hat{a}_p$ and $\hat{a}_p^\dag$ are annihilation and creation
operators, and the mode function $s_p(\eta)$ obeys the oscillatory
equation
\be
\label{eq:18}
s''_p+\omega^2(\eta)s_p =0\;
\ee
with time-dependent frequency
\be
\label{eq:19}
\omega^2(\eta) = p^2 +  \frac{1}{\chi} \,a^2 m_\varphi^2
-\l \frac16 - \xi \r \l 6 \frac{a''}{a} + \frac{\phi'^2}{M_P^2}
+ \frac{\sqrt{6}\,a^2}{M_P} \frac{\d V(\phi)}{\d \phi} \r\;.
\ee
While $\omega$ varies adiabatically, $|\,\omega'|/\omega^2 \ll 1$,
solution~\eqref{eq:17} corresponds to a particle-like field
configuration 
(note that adiabatic field evolution is a necessary condition for introducing
the very idea of particles). At this stage the particle creation does not
occur. The number density of dark matter particles is given by
\be
\label{eq:20}
n_\varphi = \frac{1}{(2 \pi a)^3} \int d^3 p \,|\beta_p|^2\;,
\ee
where $\beta_p$ is the Bogoliubov's transformation coefficient,
which relates initial and final particle-like states of the quantum field.
It can be expressed in terms of the mode functions as
\be
\label{eq:21}
|\beta_p|^2 = \frac{|s'_p|^2+\omega^2|s_p|^2}{2 \omega} -\half\;.
\ee
Assuming initially no dark matter in the Universe 
we impose the vacuum initial conditions
\be
\label{eq:22}
s_p \to 1/\sqrt{2 \omega}\,,~~~
s'_p \to -i \omega s_p\,,~~~\text{at}~~~\eta \to - \infty\;.
\ee

As mentioned above, the particle number is conserved during adiabatic
evolution. On the contrary, highly varying frequency gives rise to
particle production and the higher the variation rate the larger the
number of produced particles. In our case the maximal violation of the
adiabaticity occurs right at the end of inflation due to the
scalaron-dependent mass term. Particle creation in the case of large
$m_\varphi$ can be especially sensitive to scalaron dynamics at this
stage. Therefore we resort to numerical methods in order to find exact
evolutions of the scalaron field and the scale factor, and further
solve numerically Eqs.~\eqref{eq:18},~\eqref{eq:19} for the mode
function. Because of the limited capability of numerical calculations
we impose the initial conditions~\eqref{eq:22} not at the past
infinity but at some finite value of $\eta$. We found that the results
become practically independent of the choice of the initial time
corresponding to more than $7-8$ e-foldings till the end of inflation.
It was also learned from numerical simulations that the number of
particles in comoving volume produced at the end of inflation remains
almost constant during subsequent evolution. Indeed, the number of
produced particles is very small, hence no any Bose-enhancement is
expected during the subsequent evolution. On the other hand, both
inflaton velocity and the Hubble parameter slow down as the Universe
evolves at the post inflationary matter-dominated stage. As one can
see from~\eqref{eq:19}, this leads to strengthening of adiabaticity and
very low rate of particle production after the end of inflation.

\begin{figure}
\centerline{\includegraphics[width=0.7\textwidth]{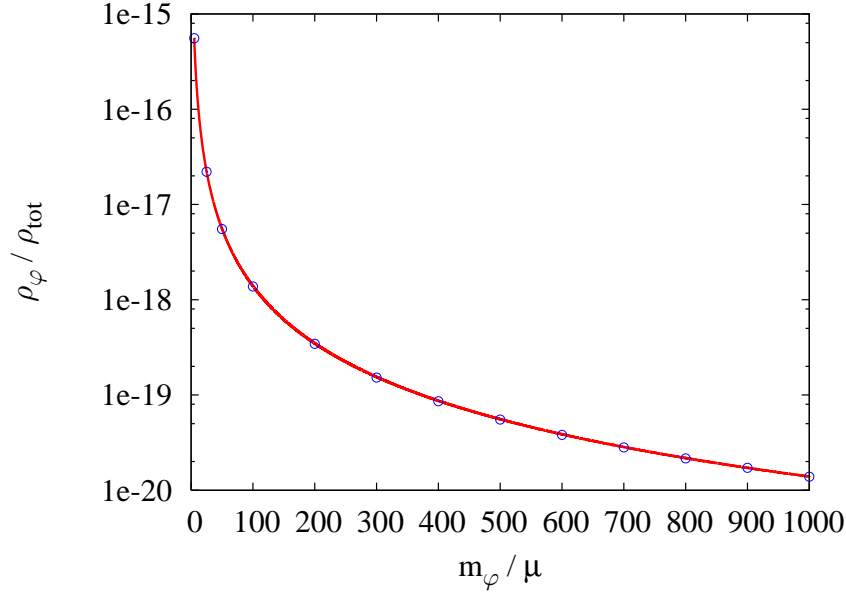}}
\caption{\label{fig:1} The ratio of the energy density of produced
scalar particles minimally coupled to gravity to the total energy
density. Small circles represent numerical results, while the line
corresponds to numerical fit with the formula~\eqref{eq:23}.}
\end{figure}

A convenient quantity for subsequent computation is the ratio of the
energy density of produced scalar particles to the total energy
density. This ratio remains almost constant while the Universe expands
due to oscillating scalaron. Numerical results for this ratio
in the case of minimally coupled scalars, $\xi=0$, are presented in
Fig.~\ref{fig:1} (small circles). It turns out that numerical data are
well described by the relation
\be
\label{eq:23}
\frac{\rho_\varphi}{\rho_{tot}}  = \alpha(\xi) \l \frac{\mu}{m_\varphi} \r^2\;,
\ee
which is plotted in Fig.~\ref{fig:1}. Here the value of
$\alpha$ for a given $\xi$ we obtained by fitting numerical data
with the formula~\eqref{eq:23}. We found that
$\alpha$ varies from $10^{-14}$ to $10^{-16}$, when $\xi$ varies from
0 to $1/4$. Note, that the resulting number
density of produced particles exhibits power law mass-dependence.
This differs from the cases of superheavy particle production 
by pure gravity (see, e.g.~\cite{Mamaev,Kuzmin:1998kk}) or by
combined effects of gravity and inflaton field with polynomial
coupling to the scalars 
(see, e.g.~\cite{Chung:1998bt}), where the mass dependence of
particles number density is exponential as mass exceeds expansion
rate. In our case at inflationary
stage the effective mass of scalars determined by the inflaton field
is very small: i.e. they are not superheavy. Number density of
produced at this stage scalars are exponentially decreases due to 
inflation. Scalars
become superheavy at $\phi \lesssim M_P$ when inflation terminates. 
The scalar mass depends exponentially on the value of scalaron field
(see Eq.~\eqref{eq:5}). Consequently the particle 
production is saturated at the end of inflation, when
scalaron-dependent mass rapidly changes from $m_\varphi \ll H$ to 
$m_\varphi > H$ violating adiabaticity,
and exponential decrease in the number density due to
the Universe expansion does not occur. From this chain of 
reasonings we conclude that it is exponentially-changing 
mass term which is responsible for the power law mass 
dependence~\eqref{eq:23}. It would be worth, however, to find more 
rigorous arguments (analytic calculations) to support the 
numerical result~\eqref{eq:23}.

Making use of Eq.~\eqref{eq:23} and the requirement that produced
scalar particles should explain dark matter phenomenon, one gets the
relation between the dark matter mass and the nonminimal coupling constant,
\be
\label{eq:24}
m_\varphi = \mu \l \frac32 \alpha(\xi)
\frac{s_0 T_{reh}}{\Omega_{DM} \rho_c} \r^{1/2}\;.
\ee
The lighter particles are overproduced in this model in accordance
with Eqs.~\eqref{eq:23},~\eqref{eq:24}. However, for very light
scalars, $m_\varphi <\mu/2$, Eq.~\eqref{eq:23} is not applicable. We
find numerically in this case that the scalar
production rate actually goes down with decreasing $m_\varphi$
so that overproduction of particles with mass of two or three orders
of magnitude lighter than the scalaron does not occur. However in this
region of masses the dominant process is
the perturbative production at reheating stage
given by Eq.\,\eqref{eq:6}.

Our final results for possible masses of scalar dark matter
given by Eqs.~\eqref{eq:14}, \eqref{eq:24} are shown in
Fig.~\ref{fig:2} by solid thick lines. The top of the figure
corresponding to superheavy dark matter particles with $m_\varphi \ge \mu/2$ is
plotted in linear scale. Here the $\xi$-dependence of dark matter
mass is almost linear as one can see from the figure. The filled
area above the curve represents cosmologically
allowed masses for any free scalars
in this model, which do not overclose the Universe.

The case of light dark matter which masses are given by
Eq.~\eqref{eq:14} is presented
in logarithmic scale. The zoom shows the fine structure of the curve
near conformal point $\xi = 1/6$. The filled region under the curve
represents allowed masses of 
free scalar particles in the model.
Allowed region is bounded from above by dashed line corresponding
to scalar mass of $10^9$\,GeV. This bound appears when we consider all possible
mechanisms of particle production which are relevant for this masses.
Indeed, from Fig.\,\ref{fig:2} one concludes that the cosmologically
allowed free scalars
with $m_\varphi \lesssim 10^{12}$\,GeV are only those which are almost conformal.
On the other hand, the
scalar particles conformally coupled to gravity are produced in
expanding Universe from vacuum fluctuations and overclose the Universe
if $m_\varphi \gtrsim 10^9$\,GeV\,\cite{Kuzmin:1998kk}.
The mass range $10^{12}~\text{GeV}
\lesssim m_\varphi \lesssim \mu/2$ is excluded, if we take into account
particle production during inflation, see the discussion
below Eq.~\eqref{eq:24}. Note, that limits obtained in
\cite{Kuzmin:1998kk} are not applicable to the model with nonminimal
coupling outside the conformal window $\xi\approx 1/6$ and scalar mass
approaching $\mu/2$ from below, see Fig.\,\ref{fig:2}, 
which we do not study in this work. 
Also, light free scalars of $m_\varphi\ll 1$\,keV may be
forbidden for $\xi$ in particular regions, where the
exponential amplification of scalar production is expected 
due to coherent effects, see Eq.\,\eqref{hot} and discussion therein.

The special choices of dark matter parameters are
  marked  by numbers in Fig.\,\ref{fig:2}. 
The points with label ``1'' correspond to warm dark
matter with mass given by Eq.~\eqref{eq:15}. The point ``2''
represents conformally coupled scalars with $m_\varphi \simeq 2.8
\times 10^{15}$~GeV. The free dark matter scalars have $m_\varphi \simeq 1.3
\times 10^{16}$~GeV, which is marked by ``3'' in Fig.~\ref{fig:2}.

Note in passing, that free superheavy fermions are also produced during
inflation. We calculate their number density as we did for scalars. 
In order to be dark matter these fermions should have mass
$m_\psi \simeq 3.1 \times 10^{15}$ GeV, which is similar to the case
of conformally coupled scalars presented above. It is worth to
emphasize, that superheavy relics produced at the end of inflation, if
unstable, cannot help to reheat the Universe earlier. As we have seen
from Fig.~\ref{fig:1}, these particles take away a very small fraction
of the energy. One can also evaluate their number density to entropy
ratio. For example, for the fermions of $10^{12}$ GeV at reheating we
have $n_\psi/s \sim \alpha \sim 10^{-15}$.  This value is much smaller
than $\eta_B$. Therefore, seesaw sterile neutrinos produced by this mechanism
cannot help with leptogenesis and one must exploit the perturbative 
production of seesaw sterile neutrinos by the scalaron
decay\,\cite{Gorbunov:2010bn}, to explain the neutrino oscillations
and the baryon asymmetry of the Universe.

To conclude, we have shown that in $R^2$-inflation the scalar particles
can form dark matter, which may be warm, cold or supercold
depending on scalar mass and the value of nonminimal coupling constant 
to gravity (see Fig.~\ref{fig:2}).

We end up with several comments. First of all, note that introduction
of selfinteracting terms for the scalars can change our predictions
for dark matter masses. Indeed, for the scattering processes which
preserve the particles number, the average momentum is conserved and
our predictions remain unchanged. However, if selfinteraction is
strong enough so that scattering to multi-particle final states becomes
rapid,
the particle number density increases while average momentum goes
down. In order to keep constant the relative contribution of the dark
matter to the present energy density, one should decrease the mass of the
candidate for a given value of $\xi$. This decreasing factor for
the mass is equal to the decreasing factor for the average momentum so that
the average velocity at the epoch of equilibrium between radiation
and matter densities would be the same. This in particular means
that the boundary values of nonminimal coupling constant (marked by 
``1'' in Fig.\,\ref{fig:2}) will  
refer to the warm dark matter case as before.

Second, superheavy dark matter, if slowly decaying, can show up in
ultra-high energy cosmic rays. 
The cosmologically long but finite lifetime can be explained by a
nonperturbative mechanisms of their decays, such as in the instanton
scenarios~\cite{Kuzmin:1997cm}, or involving quantum gravity (wormhole)
effects~\cite{Berezinsky:1997hy}.

We thank A. Khmelnitsky, I. Tkachev and A. Tokareva 
for valuable discussions.  The work is supported in part by the grant of
the President of the Russian Federation NS-5590.2012.2 and
by SCOPES program.  The work of D.G.  is supported by the Russian
Foundation for Basic Research (grants 11-02-92108, 11-02-01528). The work of
A.P. is supported by FAE program (government contract 16.740.11.0583).


\end{document}